# Exploring Estonia's Open Government Data Development as a Journey towards Excellence: Unveiling the Progress of Local Governments in Open Data Provision

Katrin Rajamäe-Soosaar, University of Tartu, Estonia
Anastasija Nikiforova[1], University of Tartu, Institute of Computer Science, Estonia

**Abstract**: Estonia has a global reputation of a "digital state" or "e-country". However, despite the success in digital governance, the country has faced challenges in the realm of Open Government Data (OGD) area, with significant advancements in its OGD ecosystem, as reflected in various open data rankings from 2020 and onwards, in the recent years being recognized among "trend-setters". This paper aims to explore the evolution and positioning of Estonia's OGD development, encompassing national and local levels, through an integrated analysis of various indices, primary data from the Estonian OGD portal, and a thorough literature review. The research shows that Estonia has made progress in the national level open data ecosystem, primarily due to improvements in the OGD portal usability and legislation amendments. However, the local level is not as developed, with local governments lagging behind in OGD provision. The literature review highlights the lack of previous research focusing on Estonian and European local open data, emphasizing the need for future studies to explore the barriers and enablers of municipal OGD. This study contributes to a nuanced understanding of Estonia's dynamic journey in the OGD landscape, shedding light on both achievements and areas warranting further attention for establishing a sustainable open data ecosystem.

## 1. Introduction

---

[1] nikiforova.anastasija@gmail.com

In acknowledging Estonia's global reputation as a "digital state" or "e-country", it is crucial to recognize a paradoxical challenge within the domain of Open Government Data (OGD). Despite its considerable success in digital government, Estonia has encountered difficulties in the OGD area, encompassing opening and maintaining these data [3], which are considered to have social, economic and environmental value for different institutions and sectors [1]. Noteworthy advancements have, however, been observed from 2020 onwards, as evidenced by Estonia's notable progress in the European Open Data Maturity Report [4]. The country was ranked 24$^{th}$ in 2018 and jumped to 5$^{th}$ position in 2020 and 2021, maintaining its leadership in subsequent years.

Despite the prominence of the European Open Data Maturity report as a widely used benchmark in Europe, it is not the only ranking that analyses countries' OGD efforts. Over the past decade, various benchmarks, including the Open Data Inventory, Open Data Readiness Assessment, Open Data Barometer, Global Open Data Index, World Justice Project Open Government Index, OGD Report and Open Data Maturity in Europe [2, 19], have been developed for ranking and comparing countries. These benchmarks utilize different methodologies and indicators, which evolve over time [19]. Consequently, an analysis of multiple ranking can provide deeper insights into the authenticity and limitations of observed advancements. As such, relying on a single report may inadvertently halt or slow down the progress of OGD initiatives by limiting a holistic understanding of the complex and multifaceted nature of open data ecosystems, influenced by diverse factors and contexts [7]. Therefore, cross-index comparison becomes crucial to gaining a more comprehensive understanding of progress and potential weaknesses. Recognizing the limitations of indices due to their broad scope, especially when covering multiple countries, a country-specific analysis may be more insightful.

To address these considerations, we conduct a systematic literature review (SLR) to explore previous research on the Estonian OGD landscape. While secondary data from indices offer valuable insights, its limitation in providing detailed causes and reasons for the results is acknowledged. To overcome this, primary data from the examination of the Estonian OGD portal is utilized. Informed by weaknesses identified in the above steps, we expand the SLR scope to include research on the O(G)D ecosystem, with a specific focus on OGD at the local or regional levels within both Estonia and Europe. Additionally, the review aims to establish a comparative analysis between findings from research on the open data ecosystem in Estonia and studies conducted on a European scale, aiming to discern similarities, differences, and potential implications for both Estonian and broader European contexts.

As such, this study investigates the evolution and positioning of Estonia's OGD development within open data rankings across various years, with a specific emphasis on the OGD provision at the local government level, since (European) cities are not not utilizing the full potential of data that are generated by them and stakeholders of this city data ecosystem [48,49]. The research findings indicate that Estonia has experienced significant upward shifts in numerous open data rankings, reflecting progress at the national level within the open data ecosystem. This notable advancement can be attributed primarily to enhancements in the usability of the OGD portal and amendments in legislation. However, a critical observation emerges when assessing the local level of OGD development, where local governments in Estonia appear to be trailing, indicating a less developed landscape compared to the national level. Despite the overall progress, there seems to be a distinct gap in the development of OGD initiatives at the local level. A comprehensive literature review focusing on the OGD ecosystem at both the local and regional levels in Estonia and Europe underscores a notable gap in previous research, particularly concerning local open data. This highlights a need for future studies to delve into the barriers and enablers specific to municipal-level OGD.

The objective of this study is twofold: 1) to examine the evolution and positioning of Estonia's OGD development, encompassing national and local levels, 2) to systematically examine and synthesize existing research literature on the open (government) data ecosystem at both the local and regional levels in Estonia and Europe.

The paper is structured as follows: Section 2 provides a methodology, Section 3 explores Estonian progress in OGD rankings over time, Section 4 examines Estonian OGD portal, Section 5 presents results of the SLR, Section 6 concludes the paper identifying future research directions.

## 2. Methodology

To achieve the study's objective, we employ a multi-faceted methodology comprising the following steps: (1) we analyze the historical performance of Estonian OGD through various rankings; (2) we examine Estonian OGD portal, which encompasses aspects such as data provision, overall architecture, and legislative influences shaping the Estonian open data ecosystem, (3) we conduct a systematic literature review (SLR) to collect and analyze relevant research on the Estonian OGD ecosystem, which, in turn, includes (3a) analysis of research specific to Estonian OGD, (3b) local level of Estonian OGD ecosystem, (3c) local/city/municipality OGD ecosystems at European level. The latter two elements of the SLR are of particular interest due to the identification of limited local OGD provision during the examination of the Estonian OGD portal. Our interest extends to assessing the advancement (if any) at the local level, which, when found limited, expanded to the European level. This expansion aims to fuel future research within the Estonian context, identifying prominent research areas and associated findings. Notably, this includes uncovering potential barriers that require exploration and overcoming in future research endeavors.

To fulfill the first step of examining the progress of Estonian OGD in rankings and indices over the years, we adopt a systematic approach. Drawing from a curated list of prominent and extensively discussed indices, including the Open Data Inventory, Open Data Readiness Assessment, Open Data Barometer, Global Open Data Index, World Justice Project Open Government Index, OECD OGD Report and Open Data Maturity in Europe [2. 19], we identify those that cover Estonia, namely (1) Open Data Inventory (Open Data Watch), (2) Open Government Index (World Justice Project (WJP)), (3) OECD Open, Useful and Re-usable data (OURdata) Index, (4) Open Data Barometer (World Wide Web foundation), (5) Open Data Maturity in Europe (European Data Portal). This involves an investigation into (1) scrutinizing the structure of each index, delving into their methodologies to identify major changes over time, (2) Estonia's performance within these indices, spanning multiple years.

To fulfill the second step, we performed an analysis of the Estonian OGD portal. This involved an examination of its content, focusing on both the diversity of topics covered and the extent of data provision by municipal and local governments.

To fulfill the third step, this study utilizes a systematic literature review (SLR) approach, to systematically examine and synthesize existing research on the open (government) data ecosystem, with a specific focus on Estonia, to identify key themes explored in prior studies, the barriers and enablers of municipal OGD, as well as to analyze the contextual factors and themes investigated in previous research addressing open (government) data at the local or regional levels within Europe. Furthermore, the review seeks to establish a comparative analysis between the findings from research on the open data ecosystem in Estonia and studies conducted on a European scale, aiming to identify similarities, differences, and potential implications for Estonian and broader European contexts. To this end, we followed the methodology defined by Kitchenham [20]. As such, the SLR involves the identification, selection, relevance assessment, and synthesis of relevant research studies.

In the first step, to achieve the SLR objective, the following questions were defined:

a) *What are the key themes that have been explored in prior research focusing on the open (government) data ecosystem of Estonia at the local government level?*
b) *What has been the context of previous research addressing open (government) data at the local or regional levels within Europe?*
c) *How can the findings from research on the open data ecosystem in Estonia be compared with those from studies conducted on a European scale?*

SLR was carried by searching digital libraries covered by Scopus and Web of Science (WoS). Given the limited number of studies we have identified in the course of this research for Estonian case, the search was complemented later by Google Scholar results. These databases were queried for keywords *"Open government data", "OGD", "open data", "local government", "municipal*", "cit*", "district", "region", "Estonia", "Europe"* that were combined using Boolean operators AND and OR. First, studies addressing open (government) data at the local government level in Estonia were identified (1st query in Table 2). Given the limited number of articles identified, totaling 10 articles after the deduplication process, the query was expanded to encompass the country level (2nd query in Table 2). This tripled the number of results in Web of Science and doubled in Scopus. Recognizing the broader European context, a third query was executed to identify relevant studies concerning municipal open data at the European level (3rd query in Table 2).

**Table 1. Queries**

| No | Search query |
|---|---|
| 1 | *("Open government data" OR "OGD" OR "open data") AND ("Estonia*" OR "EE" OR "EST") AND ("local government*" OR "LG*" OR "municipal*" OR "cit*" OR "district" OR "region*")* |
| 2 | *(Open government data" OR "OGD" OR "open data") AND ("Estonia*" OR "EE" OR "EST")* |
| 3 | *("Open government data" OR "OGD" OR "open data") AND ("EU" OR "Europe*" OR "European Union") AND "("local government*" OR "LG*" OR "municipal*" OR "cit*" OR "district" OR "region*")* |

These queries were applied to the article title, keywords, and abstract fields to narrow down the retrieved papers to primary studies, where the searched elements serve as the primary focus of the research. For the third query, the search was refined to include only English-language results from the last 5 years, ensuring a focus on recent studies (the search was conducted in August 2023).

The literature search using the 1st and 2nd queries yielded a total of 33 records in Web of Science and 65 in Scopus, with a combined count of 39 records after deduplication. Subsequently, the 3rd query produced 175 in Web of Science and 220 in Scopus, amounting to 190 distinct papers after duplicates were removed. Consequently, the cumulative number of selected records across all three queries reached 229.

To ensure comprehensive coverage, an additional search on Google Scholar was conducted using the keywords "open data" and "Estonia." The first 100 results were examined, resulting in the identification of 7 new records added to the initial results obtained from Web of Science and Scopus.

In the next step, the title and abstract of selected records (229) were screened, (see the PRISMA flow chart in Figure 1). To determine the study relevance, the following criteria was used (reading the title and abstract):

(1) the study places significant emphasis on open (government) data;
(2) open (government) data is addressed within the context of the local or regional level, encompassing areas such as data governance, citizen engagement, and public services, among others.

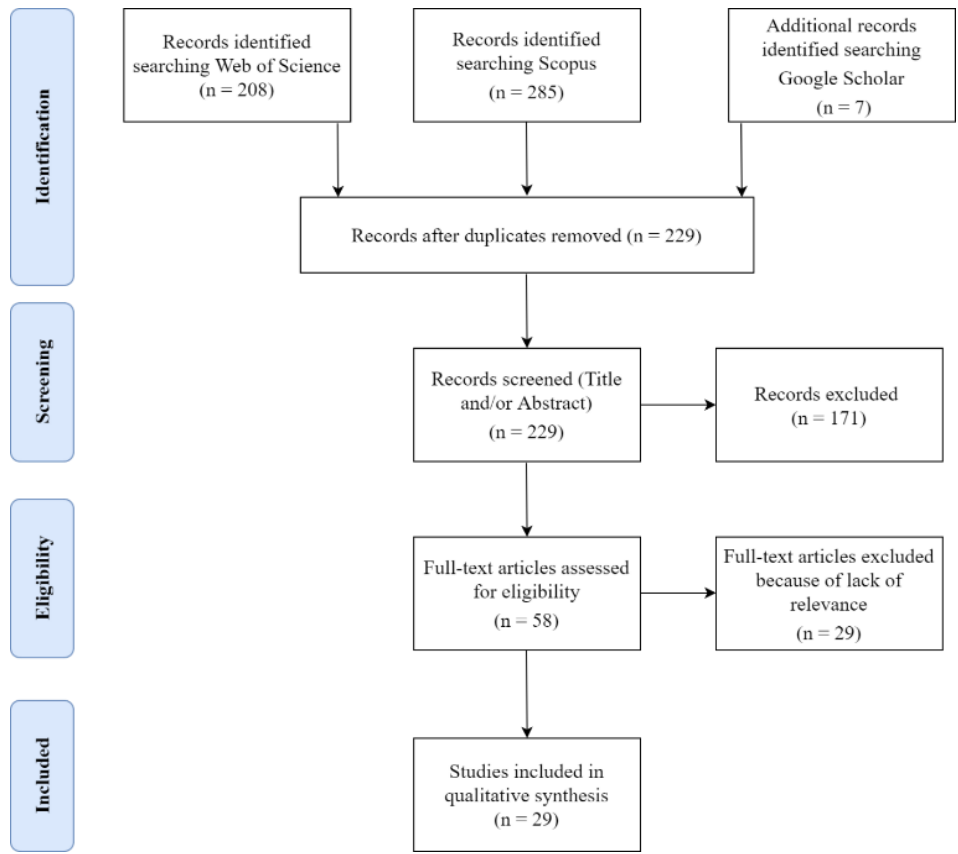

**Figure 1 PRISMA flow diagram of SLR**

This resulted in 58 papers eligible for further evaluation: 15 records addressing Q1 and 43 records for Q2. All these studies underwent a thorough assessment through the reading of their full articles. 29 papers with low relevance were excluded based on predefined criteria. Following this rigorous evaluation, the final selection for inclusion in the review comprised 29 studies that met the criteria for relevance and depth of content. To systematically analyze the selected studies, a protocol was created, where for each article the protocol extracted: (1) the descriptive data; (2) scope of the study and keywords; (3) a brief description or objectives; (4) the theory used or research method; (5) results.

The next sections present the results of the above steps.

## 3. Estonia as open data provider in EU and global rankings

In this section we analyse Estonia's OGD performance in both EU and global rankings. Our aim is to elucidate the manner and dimensions in which Estonia has evolved over time within these rankings. Through this examination, we seek to draw meaningful implications from the existing reports, shedding light on the implications and insights that emerge from Estonia's trajectory in the realm of OGD.

### 3.1. Open Data Maturity in Europe (European Data Portal)

According to European Open Data Maturity methodology [5], data for the reports spanning from 2019 to 2023 was gathered through a questionnaire distributed to national OGD representatives collaborating with the European Commission and the Public Sector Information Expert Group. The structured questionnaire aligns with four key dimensions of open data:

1. *Open Data Policy*[2] focuses on the open data policies and strategies, incorporating three indicators: (1) policy framework, (2) governance of open data, and (3) open data implementation. The dimensions underwent an update in 2022, introducing additional questions for each indicator to better account for federal and regional realities in Europe. Furthermore, there was a heightened emphasis on promoting specific data types, including geospatial data, citizen-generated data, and high-value datasets;
2. *Open Data Impact* – assesses the willingness, preparedness, and ability to measure the reuse and impact of open data. The first indicator includes strategic awareness measuring the level of reuse and impact. The second indicator was added in 2022 to gauge whether and how countries measure the reuse of open data and the methods employed. Other indicators evaluate impact within the four impact areas: the governmental (before 2022 was political), societal, environmental, and economic areas;
3. *Open Data Portal* evaluates portal features, functions provided for users, and the usage of the portal (e.g., analytics tools, responsiveness, API usage). It also delves into data provision, including local or regional data sources and portal sustainability;
4. *Open Data Quality* focuses on metadata currency (up to date'ness) and data completeness. It also monitors compliance with the Data Catalogue Vocabulary Application Profile (DCAT-AP) metadata standard, and the quality of published data deployment.

Estonia's standing in these dimensions evolved significantly, moving from 27th in 2018 to 13th in 2019 as a "follower" and leaping to 5th place within a year. Since 2020, the country has consistently held the ranking of a "trendsetter," with the best result achieved in 2023 with a notable score of 96%, compared to 93% in 2022 and 94% in 2021, as reported in the recent Open Data Maturity Report edition [6]. Figure 2 illustrates the Estonia's progress across ODM report editions, capturing the evolution in both maturity level and ranking, encapsulating the performance in each individual dimension. It showcases that the most substantial improvements occurred in the (a) impact dimension, specifically in as awareness, reuse measurement, (b) portal functionality, and (c) data quality. including monitoring and compliance.

---

[2] https://data.europa.eu/sites/default/files/method-paper_insights-report_n7_2022_0.pdf

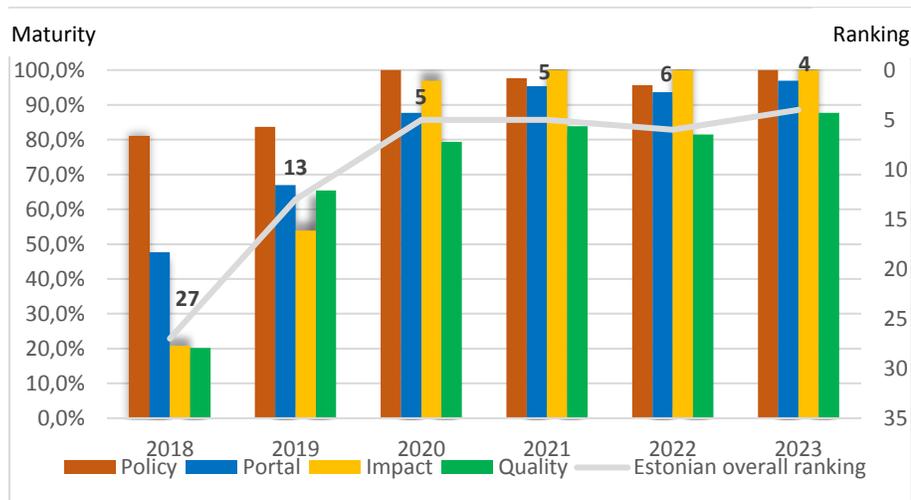

**Figure 2.** Estonian scores and ranking in European Open Data Maturity assessment 2018-2023

The maturity rating methodology underwent a revision in 2019, so the dimensions of the ranking are more comparable since 2019 based on the Report statement on changes in methodologies. enhancing the comparability of dimensions in the ranking since that year, as indicated by the Report's statement on changes in methodologies. However, despite this refinement, conducting a seamless comparison across the years remains challenging due to ongoing adjustments in methodology dimensions.

A granular examination of the dimensions reveals that *policy* indicators have consistently scored over 80% since 2018. In contrast, factors such as portal features, data provision, and portal sustainability were prominent contributors to lower scores in 2018 and 2019. While all other *portal* indicators consistently achieved scores between 95-100%, data provision data provision was still lower (72%) in 2022. Exploring the detailed country questionnaire reveals that one of the reasons is, that only bigger local governments, with more than one third of Estonia's population, publish data in machine readable formats and smaller local governments lack the knowledge and funds for it.

Examining the detailed scores within the *impact dimension* reveals notable advancements, particularly in social and economic impacts, both of which have experienced significant improvements, stabilizing at a commendable 100% for the past three years. Conversely, open data quality emerges as a challenging dimension for Estonia.

Within *open data quality*, substantial enhancements are evident in DCAT-AP compliance, surging from 53% to a 100%, and monitoring and measures, escalating from 81% to 100%. However, the indicator for data and metadata currency, including completeness since 2020, hovers around 70%. A significant contributing factor to this lower score is the predominant manual addition and editing of metadata in the national open data portal, rather than automated sourcing from the origin.

Another indicator impacting the quality dimension is deployment quality and linked data, with the latter aspect being introduced to the indicator in 2020. There has been a minor progress during 2021-2023, from 67% to 72%. The country questionnaire indicates that licencing part has been organised very well, but the linked data is a rather new concept.

Although Open Data Maturity is carried out annually, the progress assessment is difficult due to the changes in methodology, also due to the reliance on self-assessment that introduces subjectivity, with provided data contingent on the perspectives of the country representatives giving the feedback [7]. This inherent subjectivity underscores the complexity of progress assessment within the context of open data maturity.

### 3.2. OECD Open, Useful and Re-usable data (OURdata) Index

The OURdata Index benchmarks the design and implementation of open data policies at the central level, emphasizing the sustained political and policy relevance in this domain [11]. The most recent policy paper was published in 2020, data collection for which and analysis across OECD member and partner countries took place in 2018-2019. As of the end of 2023, the report has been archived. The index is structured around three pillars:

- *data availability*, which covers OGD policy framework, stakeholder engagement for data release and datasets available on central open data portal;
- *data accessibility*, which includes formal requirements (open licence, metadata), stakeholder engagement for data quality and completeness, actual implementation of these requirements;
- *government support for data reuse*, which assesses promotion of data re-use by government, value co-creation initiatives and partnerships, monitoring impact.

According to the report, Estonia received a score of 0,51, securing the $24^{th}$ position out of 32 countries. Estonia performed better in the data accessibility pillar, with the score of 0,80 ($7^{th}$ rank). Conversely, in the Government Support for Data Reuse pillar, Estonia had a lower score of 0.31. While the index has not been revised since the 2020 report, it is anticipated that Estonia's position may significantly improve, considering that the index considers similar aspects as the European Open Data Maturity (publication of OGD, policy framework, open data portal). However, it is crucial to note that the variables measured by OGD benchmarks can differ substantially, emphasizing the multidimensional and multifaceted nature of the concept of open data[2].

### 3.3. Open Data Barometer – World Wide Web foundation

The Open Data Barometer aims to unveil the prevalence and impact of open data initiatives globally, providing comparative data on governments and regions. The methodology employed combines contextual data, technical assessments, and secondary indicators [12]. The last Global version was published in 2017 and the Leaders Edition in 2018. The Barometer is structured around three sub-indexes:

- *readiness* assesses government policies and action, entrepreneurs and business, citizens and civil society;
- *implementation* evaluates the accountability dataset cluster, innovation dataset cluster and social policy dataset cluster;
- *impacts* – political, economic and social [13].

In 2016, Estonia garnered a score of 36, securing the 44th position out of 115 countries. The readiness sub-index revealed that citizens and civil rights received the highest score of 81, while entrepreneurs and business - the lowest score of 31. However, the impact scores were notably low, registering at 0 in the political and social dimensions and 20 in the economic dimension.

It's noteworthy that Estonia's lower scores align with the findings of the OECD OURdata Index. This consistency can be attributed to the lesser interest in open data from policymakers until 2018 when OGD provision gained increased attention. The alignment in scores suggests a parallel evolution in the recognition and prioritization of open data initiatives within Estonia's policy landscape

### 3.4. Open Data Inventory (Open Data Watch)

The Open Data Inventory (ODIN) serves as an assessment tool for evaluating the coverage and openness of data available on the websites maintained by national statistical offices (NSOs) and any official government website accessible from the NSO site [8]. In the case of Estonia, the evaluation is based on the website www.stat.ee/en/node, which is overseen by the national agency Statistics Estonia. In other words, it does not cover OGD initiative as a whole.

The Estonian profile within the Open Data Inventory (ODIN) framework, shown in Figure 3, highlights certain areas requiring attention, with coverage identified as a key aspect in need of improvement [9]. Notably, challenges persist in the social statistics subscore and the availability of data on the subnational level that have led to lower scores. Despite these challenges, the overall score ranks Estonia 5$^{th}$ in Northern Europe and the 11$^{th}$ position globally in 2022.

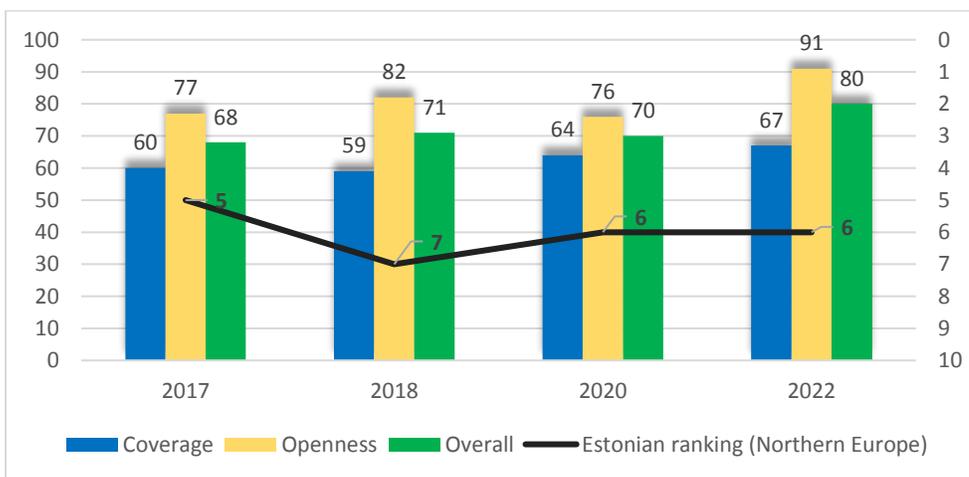

**Figure 3.** Estonian score and ranking of ODIN (2017-2022)

### 3.5. Open Government Index - World Justice Project (WJP)

While the World Justice Project is not primarily focused on OGD, Open Government plays a pivotal role as one of the essential components within Factor #3 of the Index. The Open Government Index evaluates the extent to which basic laws and information on legal rights are publicized, in addition to scrutinizing the quality of information disseminated by the government [10]. It comprises four dimensions:
- *publicized laws and government data* – the public availability of basic laws, government information, drafts of legislation etc.;
- *right to information* – the process of requesting information from government agencies, considering completeness, response time, and associated costs;

- *civic participation* – protection of the freedom of opinion, assembly, and the right to petition, along with the availability of information regarding decisions impacting the community;
- *complaint mechanisms* – the ability to file complaints about the provision of public services or the conduct of government officers in fulfilling their legal duties.

Estonia's score for 2023 stands at 0.81, positioning the country at the 9th place globally and the 7th in the regional ranking. Estonia has maintained a score of at least 0,80 since 2019. Notably, within the sub-factors, the highest and most important score from the OGD perspective is of publicized laws and government data (0.88), with 5th position in the global ranking. However, sub-scores for "Civic Participation" and "Complaint Mechanisms" have been comparatively weaker, resulting in the 13th position out of 31 countries in the regional ranking [10].

### 3.6. Discussion

In the summer of 2018, the national open data portal in Estonia demonstrated limited data availability, providing only 89 datasets [3]. Data holders were required to create an open data catalogue on their website and forms for requesting open data, coupled with intricate procedures for users to request open data, leading to inconvenience and delays [14].

Unlike many European countries, Estonia boasted the X-Road - a national interoperability infrastructure, often cited in literature as a reason for the subdued focus on open data within Estonia [3]. The X-Road provides unified and secure data exchange, extensively utilized by Estonian public sector institutions for data exchange and citizen service delivery. As it enables data reuse also for private organisations, as a significant portion of required data can be sourced from X-Road services. The X-Road was probably the main cause why Estonian policymakers remained sceptical of OGD initiatives, although there was societal and international pressure, including the importance of OGD in international e-Government rankings [15] [14].

In 2018, the Estonian Ministry of Economic Affairs and Communications entered into a contract with Open Knowledge Estonia (non-profit organization) to improve Estonia's OGD ecosystem. According to [14], by 2020 substantial improvements in various performance indicators was identified: (1) number of datasets; (2) unique users; (3) number of applications on portal; (4) OGD events and news articles and social media groups; (5) ranking of Estonia in European OD Maturity report jumped from 27$^{th}$ to 14$^{th}$ place.

Legislation played a pivotal role in this transformation. By the end of 2021, amendments to the Estonian Public Information Act entered into force to bring the Estonian legislation in line with EU Directive 2019/1024 on open data. The amendments introduced several obligations related to public information provision and re-use, including [16]:

(1) open data should include data sets and data descriptions;
(2) data that are collected or produced in the course of scientific research activities and are used as evidence in the research process, or are commonly accepted in the research community as necessary to validate research findings and results (research data), shall be made available for re-use, if the production of research data has been funded from the budget of the state, local governments or legal persons in public law and researchers, research performing persons or research funding persons have already made them publicly available through an institutional or subject-based repository. Scientific publications shall not be deemed to be research data;

(3) open data that are subject to frequent or real-time updates (dynamic data) shall be made available for re-use immediately after collection, or in the case of a manual update immediately after the modification of the data, via an application programming interface (API) and, where relevant, as a bulk download;

(4) the re-use of open data shall generally not subject to any conditions. If imposing conditions for making the data available for re-use are necessary in the public interest, such conditions shall be objective, proportionate and non-discriminatory;

(5) open data, the re-use of which is associated with important benefits for society, the environment, and the economy (high-value datasets), should be made available for re-use by the holder of the information free of charge in machine-readable format via a suitable API and, where relevant, as a bulk download;

(6) income from supplying information for re-use shall not exceed the costs incurred for the reproduction, provision, and dissemination of open data as well as for anonymization of personal data and protection of business secrets.

The amendments introduced several obligations related to public information provision and re-use, emphasizing aspects such as open data descriptions, availability of research data, immediate release of dynamic data, non-restrictive re-use conditions, and free-of-charge access to high-value datasets.

Although the Public Information Act encompasses a broader concept of public information, its relevance to open data provision is paramount, as OGD is part of public information.

These legislative strides, coupled with improvements in open data portal usability, emerged as crucial factors contributing to Estonia's rise in rankings in both the European Open Data Maturity and WJP Open Government Index. However. while there has been substantial progress at the national level, the local level presents challenges. The European Open Data Maturity questionnaire indicates that only larger local governments, housing over one-third of Estonia's population, publish data in machine-readable formats, with smaller local governments lacking the necessary knowledge and funds.

This is also in line with [2] that compared open data benchmarks in 2021, concluding that most benchmarks primarily focus on governments, especially at the national level, with only one, the Global Open Data Index, including a focus on regional or local levels. Unfortunately, Estonia was not included in the Global Open Data Index, and the project has been archived, limiting in-depth insights into the country's local open data landscape.

The next section will provide an overview of the current status of the Estonian OGD portal and municipal data provision by analyzing Estonian open data portal.

## 4. Estonian Open Data Portal
### 4.1. General portal profile

Estonian launched its first national open data portal in 2015, built on the CKAN platform - one of the most popular open-source data management systems adopted for OGD portals [17][18]. However, the initial implementation had limitations, with many features of the CKAN platform left unused, and the portal hosted a relatively small number of datasets [14].

In 2020, the Estonian Ministry of Economics and Communication initiated an analysis and development process for a new open data portal. Based on the documentation of the Estonian Open Data Portal [19], the Ministry initially planned to utilize the CKAN open-source platform. However, after conducting a thorough analysis, the decision was made to abandon this initial plan. The rationale presented in the document highlighted that the required functionalities would have necessitated a significant redevelopment of most of the existing CKAN components. Consequently, the ministry opted for an alternative approach and proceeded to develop a custom-made platform to better meet the specific needs and requirements of the new open data portal.

The current open data portal, avaandmed.eesti.ee, was officially launched in 2021. As of January 14, 2024, the Estonian open data portal provides 1807 datasets - the number more than tripled ,compared to 582 datasets in May 2020 [14]. The portal lists 2232 publishers, including public sector institutions, some of which are yet to publish any data.

The portal has the highest number of datasets in categories such as education, culture and sport (493 datasets); population and society (457); and science and technology (430). Conversely, categories like energy and agriculture, fisheries, forestry, and food have a limited number of datasets, with only 10 and 53, respectively (see Figure 4).

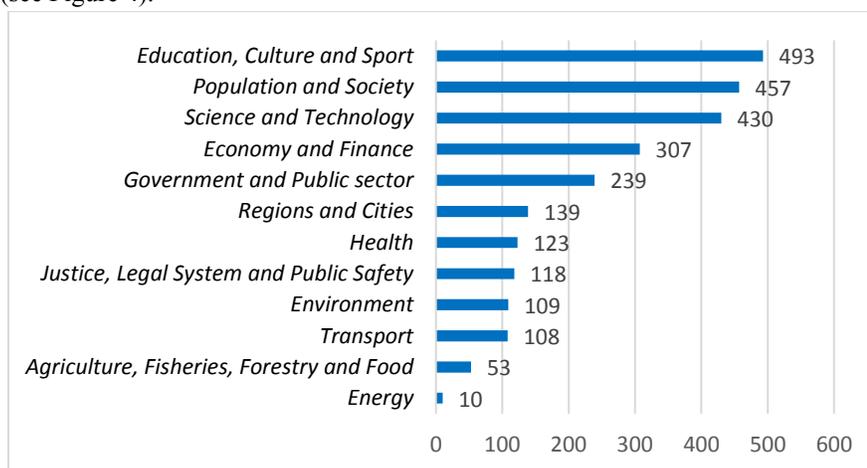

**Figure 4.** Number of datasets per category in avaandmed.eesti.ee (14.01.2024)

### 4.2. Municipal open data provision

From the perspective of this study, which is interested in local government data, the most relevant category on the national portal is "Regions and Cities" with 139 datasets. A more detailed examination unveils that the primary publishers within this category include the City Governments of Tallinn and Tartu, along with the Estonian Land Board. Only 12 of 79 local governments (15%) have published some data on the national portal (Table 1).

**Table 1. The datasets of Estonian local governments in national portal (14.01.2024)**

| Municipality Government | Population (01.01.2023) | No of data-sets | Type of data |
|---|---|---|---|
| Tallinn | 426 538 | 102 | Geospatial data, budget, sensor data, services, transport, accessibility, statistical overviews etc |

| Tartu | 97 435 | 34 | Geospatial data, master plan, document register, contacts, transport, cemeteries, education etc |
|---|---|---|---|
| Sillamäe | 12 157 | 7 | Education, main documents, budget |
| Alutaguse | 4 678 | 6 | Population, budget, main regulations and documents |
| Rapla | 13 228 | 4 | Procurement plan, budget, document register |
| Pärnu | 51 874 | 2 | Events and activities of interest, planning |
| Valga | 15 456 | 2 | Transport, document registry |
| Narva | 53 625 | 1 | Detailed spatial planning |
| Loksa | 2 498 | 1 | Document register |
| Anija | 6 431 | 1 | Document register |
| Harku | 17 520 | 1 | Document register |
| Saue | 25 571 | 1 | Document register |

As anticipated, the largest cities in Estonia, namely Tallinn and Tartu, are the foremost contributors of data among local governments. Both cities have dedicated local portals for geospatial data, accessible through https://geohub.tartulv.ee/ and https://www.tallinn.ee/en/geoportal.

In adherence to the Estonian Public Information Act, local governments are mandated to maintain a document register and disclose certain documents, such as legislation, contracts, and public letters, from their Document Management System (DMS) [16]. The public document register contains documents that are not subject to restrictions (e.g. private data of persons). Local government websites typically feature direct links to the public view of their DMS, where, in the case of nearly 50 local governments, JSON files containing local governments' legislation, contracts, and official correspondence can be accessed and downloaded. However, only a few municipalities have published links to their DMS datasets on the Open Data portal. In the case of Anija, Harku and Saue Municipalities, the portal specifies that the data has been migrated from the previous version of the national OGD portal and has not undergone a review by the publisher. On the other hand, other local governments have updated the metadata information of their datasets on the national portal.

In summary, while Estonia has achieved notable advancements in various open data rankings, the local governments are lagging behind. To sustain Estonia's success in the OGD domain and recognize the growing importance of local and smart city open data portals within the broader open data ecosystem, it is imperative to enhance the current state of local open data. The research has identified some simple and straightforward steps for improvement, e.g. although many local governments have provided links to their DMS public view with possibility to obtain their legislation, contracts, and correspondence in machine-readable format on their official websites, these datasets are not yet included in Estonian official OGD portal.

## 5. Results of the systematic literature review

As part of the SLR, our objective was twofold: (1) to ascertain the extent of Estonia's representation in the scientific literature concerning OGD, and (2) to delineate the specific OGD topics that are explored at the local government level in Europe. The findings and summaries of these studies are presented in Tables 3 and 4.

Six out of the eight papers were published between 2018-2020, shedding light on the limited progress within Estonia's open data ecosystem, despite the country's advanced e-government infrastructure (Table 3). Among these papers, six delve into specific Estonian case studies, investigating aspects such as the reasons behind the country's low OGD maturity [3] [26]. The remaining two studies incorporate Estonia into broader case studies, exploring Northern European cross-country peer-to-peer communication [25] and conducting an exploratory case study on OGD usage within crisis management contexts [22]. It is noteworthy that none of the studies in this review specifically address the local government level in Estonia.

**Table 3. Studies addressing the first SLR objective – Estonian OGD ecosystem**

| No | Ref. | Scope | Objective of the study |
|---|---|---|---|
| 1 | [21] | Estonian case study, public service co-creation | To understand the transformation brought by the accessibility and utilization of OGD in co-creating public services, empowering service users, transforming them into active collaborators rather than customers of public service providers |
| 2 | [14] | Estonian case study, OD ecosystem | To improve the performance Estonian OGD ecosystem though the use of action research and systems theory |
| 3 | [22] | Case study of Czech Republic, Estonia and Latvia, crisis management | To understand how OGD can be used during times of crisis as a crisis management tool and how does OGD influence the co-creation of services that assist in crisis management |
| 4 | [23] | Estonian case study, civic engagement | To understand the potential of open data phenomena to promote public sector innovations and civic engagement |
| 5 | [24] | OD policies, Estonian case study | To provide an overview of existing systems to ensure access to open information and make proposals on how to improve open data disclosure practices in Estonia |
| 6 | [25] | Nordic Europe, peer to peer communication | To explore a collaborative nature of peer-to-peer interactions in the open data area and further debates on the potential of these data-driven networks and platforms to transform the classical mechanisms of co-production and public sector innovations in e-government not only between traditional open data actors, but between peers of the movement themselves |
| 7 | [3] | OGD, Estonian case study | To explore Estonia's low OGD maturity against the backdrop of a highly developed e-government by conducting an exploratory case study using document analysis, survey data and semi-structured interviews |
| 8 | [26] | Open data politics, Estonian case study | To analyse the development of open data phenomena in a country known as one of the global leaders in promoting information society, demonstrating advances in building sophisticated e-government, e-commerce, e-voting and blockchain governance ecosystems |

The topics addressed in these studies include: (a) open (government) data ecosystem maturity – stakeholders, barriers, enablers, policy proposals etc.; (b) OGD use for public service co-creation; (c) peer-to-peer perspective on open data.

Table 4 provides an overview of studies addressing open (government) data within the context of local or regional levels in Europe, encompassing themes such as data governance, local citizens, public services etc. Notably, although the search terms did not include keyword "smart cities", it is noteworthy that the predominant focus in studies concerning OGD at the local government level often revolves around the concept of "smart" cities or data.

Other selected papers address the open data of local governments in broader context, e.g. (1) data governance, data operability and linked data [34] [35] [36] [37]; (2) transparency and citizen trust [38] [39]; (3) Society 5.0 [40] [41]; 4) public administration and public services in connection with OGD [42] [43].

**Table 4. Studies addressing the context of European OGD ecosystem at regional or local level**

| No | Ref. | Scope | Objective of the study |
|---|---|---|---|
| 1 | [34] | OD and trust; European countries | To assess whether the extent of openness and the coverage of data sets released by European governments significantly influence citizen trust in public institutions |
| 2 | [35] | OGD trends | To evaluate the global progress and explore research areas and development trends of OGD field based on the SLR |
| 3 | [36] | OD re-use, relevant social groups | To clarify open data use and engagement outside the public sector to strengthen the empirical base for the understanding of open data use |
| 4 | [37] | Linked open data, EU CPSV | To develop a process for using CPSV-AP and conduct a pilot implementation, using this process, to investigate potential benefits or challenges from its use |
| 5 | [38] | LOD, European interoperability | To understand how LOD principles and technologies can be applied for the publication and interlinking of public administration reporting data |
| 6 | [39] | Fair data ecosystem, EU | To introduce the concept of fair data ecosystem as an alternative to corporate-driven, state-led, and citizen-centric approaches to digital transformation |
| 7 | [27] | OD platforms and smart cities | To investigate the relationship between smart urban development and the use of open data platforms; understand how these are useful for defining actions and strategies that facilitate the planning of a smart city, and to find platform's common characteristics that allow cooperation of intentions between European Union cities |
| 8 | [28] | OD platforms, smart cities, agency | To understand how open data platforms are coproduced by different actors based on their conceptions of open data in two European cities, Lyon and Berlin. |
| 9 | [29] | OD portal assessment, data reuse | To assess compliance of the Croatian Open Data Portal with user-oriented principles that sustainable open data portals should implement adopting the metrics proposed by the European Commission to |

| No | Ref. | Scope | Objective of the study |
|----|------|-------|------------------------|
| 10 | [40] | OD policy, citizen participation, OD reuse | To examine how does the Open Data Directive align with and diverge from the rationale and requirements of the movement for open data, and what are the implications of this for citizen participation |
| 11 | [33] | Digital maturity, urban municipalities, EC Intelligent Cities Challenge | To analyse the digital maturity self-assessment results undertaken by 11 Slovenian urban municipalities utilising ICC's assessment methodology framework for government services and social connectivity |
| 12 | [41] | Local government transparency, SLR | To investigate the extent of research interest in local government transparency from 2000 to 2018, identifying research gaps |
| 13 | [30] | Open data, data interoperability, common semantics | To enhance the data sharing processes in Italy-Switzerland cross-border area, particularly addressing tourism and mobility that are key economic activities for the region through the review on the data catalogues published in dati.lombardia.it and opendata.swiss. |
| 14 | [31] | Open data, monitoring, COVID-19 | To analyse the opportunities and critical issues surrounding the use of open data to improve the quality of life during the COVID-19 epidemic, as well as for the effective regulation of society, the participation of citizens, and their well-being |
| 15 | [42] | OD based public services, co-creation | To develop methods for co-creating open digital services for age-friendly cities and communities enabling civic open data use of older adults, increasing digital inclusion of older adults, and co-creating sustainable digital public services for older adults |
| 16 | [43] | Big and open linked data, business model canvas | To develop canvas to describe and develop business models for creating value from big and open linked data (BOLD) in smart and circular cities (SCCs) |
| 17 | [44] | OGD portal, smart data, Society 5.0 | To identify whether OGD portals in various countries support the open (government) data initiative and the movement to "smarter" open data, incl. high-value data, and whether they are suited for further reuse |
| 18 | [32] | Educational open data | To explore the role of open data in the education and the practical application of acquired knowledge. Current situation in Serbia in terms of educational open data was analysed, providing suggestions for improvement |
| 19 | [45] | Open data ecosystem, OGD portal | To assess the state of open data in Croatia via application of the assessment framework developed during the Online Training Program (OTP) of Horizon 2020 project, discussing the usefulness of such evaluations based on the interpretation of the assessment results. |

| No | Ref. | Scope | Objective of the study |
|---|---|---|---|
| 20 | [2] | Open data benchmarks, ranking | To compare the metrics and methodologies used to measure, benchmark, and rank governments' progress in OGD initiatives. Comparison between the various existing benchmarks at a single moment in time and between each benchmark at different moments in time |
| 21 | [46] | OGD, Smart City, Society 5.0 | To define the Society 5.0 and OGD concepts and emphasize their interconnection, as well as to provide real-world examples proving these concepts are interconnected |

21 papers were deemed most relevant, with seven specifically delving into the analysis of a distinct European region or municipality. The primary areas of focus in these studies are: (1) open data platform or portal [27] [28] [29], (2) usage of open data and data sharing [30] [31] [32], and (3) the assessment of digital maturity, which includes an evaluation of open data, within municipalities [33].

## 6. Conclusions

This study aimed to examine the development of Estonia's OGD progress, its positioning in open data rankings, encompassing the OGD provision by local governments. The findings indicate that while Estonia has made significant strides in ascending open data rankings - the European Open Data Maturity and WJP Open Government Index, local governments, as a whole, are trailing. The largest cities, Tallinn and Tartu, unsurprisingly dominate as major data publishers, yet 85% of municipalities haven't contributed any data to the national OGD portal. Recognizing the growing significance of local and smart city open data portals within the broader ecosystem, it is imperative to address the current limitations in local open data provision and enhance the visibility of local OGD in the national portal.

The objective of the SLR was to: (1) review existing academic literature pertaining to Estonia's open (government) data ecosystem, (2) determine the context of previous studies concerning the local and regional levels of Europe, (3) compare of results from research on Estonia's open data ecosystem with broader European perspectives. The SLR revealed that studies focused on the Estonian case, particularly those published between 2018 and 2020, underscore the slow progress of Estonia's open data ecosystem despite its advanced e-government. Notably, none of the previous studies delved into the local government level of Estonia. The primary themes identified in European studies at regional or local levels included open data portals, open data usage and sharing, digital maturity assessments of municipalities, data interoperability and linked data, transparency, citizen trust, Society 5.0, and the intersection of public administration and services with OGD.

While this study has suggested some steps to enhance OGD ecosystem in Estonia, including local government data provision through the central OGD portal, further in-depth exploration is required to identify the primary barriers preventing municipalities from openly sharing OGD. For this, future research can consider conducting semi-structured interviews with local government officials, mapping the main user groups of local-level OGD, developing detailed use-cases of data reuse. Ecosystem approach would be valuable to bring more comprehensive view effectively examining components of local ecosystem, identifying gaps and defining corrective actions [48]. As the SLR has highlighted the limited scope of regional and local-level OGD provision research, future investigations hold the promise of providing new insights into the barriers and facilitators of municipal OGD. For this, we find qualitative analysis of local governments resistance towards openly publishing data to be the next step of our research.